\documentclass[twocolumn]{aastex63}

\usepackage{units}

\usepackage{amsmath}

\newcommand{\code}[1]{\texttt{#1}}
\newcommand{\mesa}{\code{MESA}}
\newcommand{\MESA}{\mesa}

\newcommand{\Msun}{\ensuremath{\mathrm{M}_\odot}}
\newcommand{\Lsun}{\ensuremath{\mathrm{L}_\odot}}

\newcommand{\Dmix}{\ensuremath{D_{\mathrm{mix}}}}
\newcommand{\ALi}{\ensuremath{A(\rm Li)}}

\newcommand{\nuclei}[2]{\ensuremath{\mathrm{^{#1}#2}}}

\newcommand{\helium}[1][4]{\nuclei{#1}{He}}
\newcommand{\lithium}[1][7]{\nuclei{#1}{Li}}
\newcommand{\beryllium}[1][9]{\nuclei{#1}{Be}}

\begin{document}

\received{July 25, 2020}
\revised{August 23, 2020}
\accepted{August 31, 2020}

\author[0000-0002-4870-8855]{Josiah Schwab}
\affiliation{Department of Astronomy and Astrophysics, University of California, Santa Cruz, CA 95064, USA}
\correspondingauthor{Josiah Schwab}
\email{jwschwab@ucsc.edu}

\title{A helium-flash-induced mixing event can explain the lithium abundances of red clump stars}

\begin{abstract}
  Observations demonstrate that the surface abundance of \lithium[7]
  in low-mass stars changes dramatically between the tip of the red
  giant branch and the red clump.  This naturally suggests an
  association with the helium core flash, which occurs between these
  two stages.  Using stellar evolution models and a simple, ad hoc
  mixing prescription, we demonstrate that the \lithium[7] enhancement
  can be explained by a brief chemical mixing event that occurs at the
  time of the first, strongest He sub-flash.  The amount of
  \beryllium[7] already present above the H-burning shell just before
  the flash, once it mixes into the cooler envelope and undergoes an
  electron capture converting it to \lithium[7], is sufficient to
  explain the observed abundance at the red clump.  We suggest that
  the excitation of internal gravity waves by the vigorous turbulent
  convection during the flash may provide a physical mechanism that
  can induce such mixing.
\end{abstract}


\section{Introduction}

Recently, \citet{Kumar2020b} used data from GALAH DR2 \citep{GALAHDR2}
and \textit{Gaia} DR2 \citep{GaiaDR2} to demonstrate that red clump
(RC) stars have an enhancement in \lithium[7] that is a factor of
$\approx 40$ above the value near the tip of the red giant branch
(TRGB).  This measurement is enabled by asteroseismic diagnostics that
allow the separation of H-shell-burning-only red giant stars and the
He-core-burning red clump stars \citep[e.g.,][]{Mosser2011,
  Bedding2011, Hawkins2018a, Ting2018c}.  Significant surface lithium enhancement is occurring
between the TRGB and the RC, where the notable interior stellar event
that occurs in this interval is the helium core flash \citep{Thomas1967}.
In this letter, we demonstrate that mixing in the envelope triggered
by the first, strongest He sub-flash and the action of the
Cameron-Fowler process \citep{Cameron1955, Cameron1971} naturally
explains this observation.

\section{Stellar models}

\begin{figure}
  \centering
  \includegraphics[width=\columnwidth]{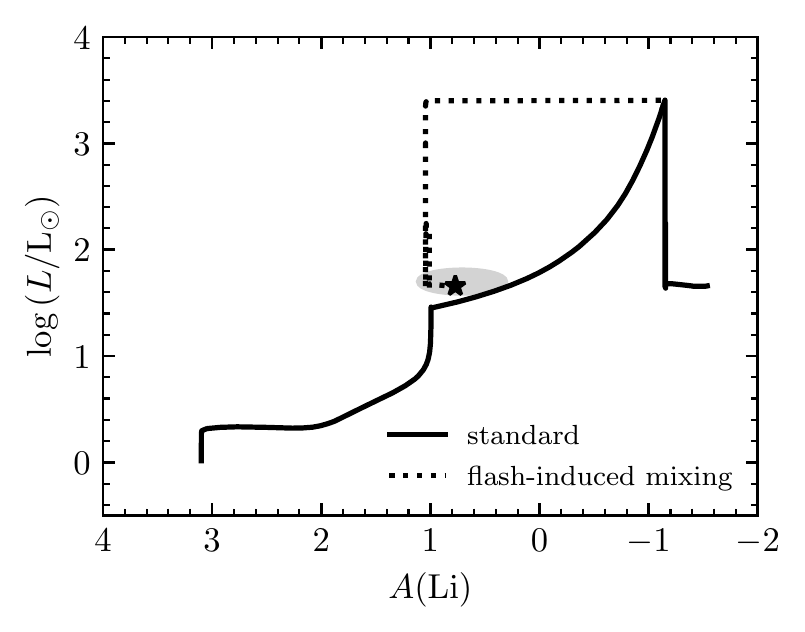}
  \caption{Evolution of \lithium[7] abundance with stellar luminosity
    in the $\unit[1]{\Msun}$ model.  The star indicates the location
    of the flash-induced mixing model when the star is on the RC.  The
    shaded ellipse indicates the location of the peak density of RC
    stars found by \citet{Kumar2020b}.}
  \label{fig:ALi}
\end{figure}

We construct stellar evolution models using \MESA\ \citep{Paxton2011,
  Paxton2013, Paxton2015, Paxton2018, Paxton2019}.  We evolve a
representative model of a \unit[1]{\Msun} star at solar metallicity.%
% using a set of standard \MESA\ options appropriate for modeling
% low-mass stars.%
\footnote{Our \MESA\ input and output files are publicly available on
  Zenodo. \url{https://doi.org/10.5281/zenodo.3960434}}
We adopt a mixing length of 1.8 times the pressure scale
  height.  Following \citet{Kumar2020b}, motivated by past work
requiring ``extra'' mixing in RGB stars
\citep[e.g.,][]{Charbonnel2007, Angelou2012}, these models include
thermohaline mixing using the \citet{Kippenhahn1980a} prescription
with an efficiency of $\alpha_{\rm thm} = 100$.  This gives the deep
mixing necessary to destroy \lithium[7] on the first ascent giant
branch.  The model is non-rotating, and while rotation may play
  a role in setting the \lithium[7] abundance on the MS and RGB \citep[e.g.,][]{Charbonnel2005, Charbonnel2010} and rapid rotation a role in producing Li-rich RGB and RC stars \citep[e.g.,][]{Martell2020}, 
  rotation seems an unlikely explanation for a ubiquitous mixing event occurring between the TRGB and RC.

We use the standard \MESA\ \texttt{pp\_and\_cno\_extras} nuclear
network, which includes 25 species and the reactions covering the
pp-chains and CNO-cycles.
The \MESA\ nuclear reaction rates are a combination of rates from
  NACRE \citep{Angulo1999} and JINA REACLIB \citep{Cyburt2010}.  The key rates $\helium[3](\alpha,\gamma)\beryllium[7]$ and $\lithium[7](p, \alpha)\helium[4]$
are drawn from NACRE below $\unit[10^7]{K}$ and REACLIB above that temperature.
We use the electron-capture rate on $\beryllium[7]$ 
from \citet{Simonucci2013}, as made available in
machine-readable form by \citet{Vescovi2019}.\footnote{The \MESA\ REACLIB  $\beryllium[7](e^-,\nu_e)\lithium[7]$ reaction rate does not correctly incorporate the electron capture from
  bound states.  This causes the rate to be dramatically underestimated at temperatures below $\unit[10^7]{K}$.}  Models are
initialized on the pre-main sequence with the \citet{Asplund2009}
solar abundance pattern and $Z = 0.014$.  This initializes lithium to
the meteoritic abundance $\ALi = 3.26$.

The solid line in Figure~\ref{fig:ALi} shows the standard evolutionary
track.  This closely matches the \MESA\ result shown as the dashed
line in Figure~3 of \citet{Kumar2020b}.  Relative to that track, there
is a horizontal shift of a few tenths of a dex in the value of
$A({\rm Li})$ as the \citet{Kumar2020b} calculation was initialized to
match the observed abundance of Li at the main sequence turn-off.
With our adopted mixing during the main sequence, lithium is not
depleted to the observed main sequence level, as some extra
mixing at the base of the convection zone is typically invoked to
explain the observed values \citep[e.g.,][]{Baraffe2017b}.%
This discrepancy is not important for our subsequent discussion
  of the abundance on the RC as the Li experiences substantial dilution
  and further destruction during the RGB.

It has long been understood \citep{Cameron1955, Cameron1971} that one
way to produce \lithium[7] is by having a reservoir of \beryllium[7],
produced via $\helium[3](\alpha, \gamma)\beryllium[7]$ at interior
temperatures $T \gtrsim \unit[10^7]{K}$, that is mixed out into a
cooler convective envelope, where the $\beryllium[7]$ undergoes an
electron capture to form $\lithium[7]$.  We posit that the He flash
initiates some mixing that activates this process.

A detailed graphical summary of a typical helium core flash in a
\unit[1]{\Msun} \MESA\ model is shown in Figure~2 of
\citet{Bildsten2012}.  The transition to core He burning occurs
through a series of sub-flashes that occur over a period of
$\approx \unit[2]{Myr}$.  The first flash is the strongest, with the
peak He-burning luminosity exceeding $\unit[10^{9}]{\Lsun}$.  Subsequent
sub-flashes have more modest peak He-burning luminosities
$\approx \unit[10^3-10^4]{\Lsun}$.
Carrying these high luminosities out from the location of the
temperature peak requires convective transport, though because the
energy goes into expansion (and hence need not be transported through
the entire star), the convective zone remains restricted to the He
core.

Internal gravity waves (IGWs) can be excited by turbulent convective
motions \citep[e.g.,][]{Press1981, Goldreich1990, Lecoanet2013}.
Recently, \citet{MillerBertolami2020} suggested that IGWs excited by
the He flash might induce photometric variability in hot subdwarf
stars (RG cores without large H envelopes) as they evolve towards the
He main sequence (RC equivalent).
Here we speculate that these IGWs may induce chemical mixing in the
region just above the H-burning shell, possibly either wave-driven
mixing or wave-driven heating that triggers convective mixing, linking
this region with the convective envelope.
We note that the effects of IGWs have been previously invoked to
explain lithium evolution in other stellar evolution contexts,
including mixing as a cause of Li depletion in F stars
\citep{GarciaLopez1991},
and the Li abundances of low mass, solar-type stars \citep{Montalban1994, Charbonnel2005}.

A larger Mach number ($\mathcal{M}$) of the convective motions implies
increased excitation of IGWs \citep{Lecoanet2013}.
During the first sub-flash, the maximum convective Mach number predicted from
mixing length theory can reach \mbox{$\mathcal{M} \sim 10^{-2}$},
while the later sub-flashes only reach Mach numbers of
\mbox{$\mathcal{M} \sim 10^{-4}$}.
Just before the flash, the characteristic luminosity near the base of
the H-rich layer, $L_{\rm H}$, is set by the H-burning shell to
$\sim 10^{3.5}\,\Lsun$ (i.e., the TRGB luminosity).  If we take the
IGW wave luminosity to be $L_{\rm IGW} \sim \mathcal{M} L_{\rm conv}$, then
for this first sub-flash $L_{\rm IGW} \gg L_{\rm H}$, while subsequent
flashes have $L_{\rm IGW} \ll L_{\rm H}$.  This motivates our suggestion
that the first sub-flash may be special in its effects.

To demonstrate that such a mixing event can explain the observed Li
enhancements, we adopt a simple, ad hoc model of flash-induced mixing.
In the H-rich envelope material $(X > 0.5)$, we apply a constant
diffusive mixing coefficient of
$\Dmix = \unit[10^{12}]{cm^2\,s^{-1}}$.  This diffusivity is additive
with any other diffusive mixing processes that are occurring (i.e.,
convective and thermohaline mixing) and the chosen value is
intermediate in magnitude between the two.  In order to tie this
mixing to the He flash, we activate it only when the helium-burning
luminosity is large, specifically when
$\log(L_{\rm He}/\rm \Lsun) > 4$, so that mixing occurs primarily
during the first He sub-flash.

The dotted curve in Figure~\ref{fig:ALi} shows the result of a \MESA\
model including this flash-induced mixing.  At the first sub-flash,
the star is enhanced up to $\ALi \approx 1$ while still near the TRGB
luminosity and then experiences only a small amount of depletion from
that value as it goes to the RC.

\begin{figure}
  \centering
  \includegraphics[width=\columnwidth]{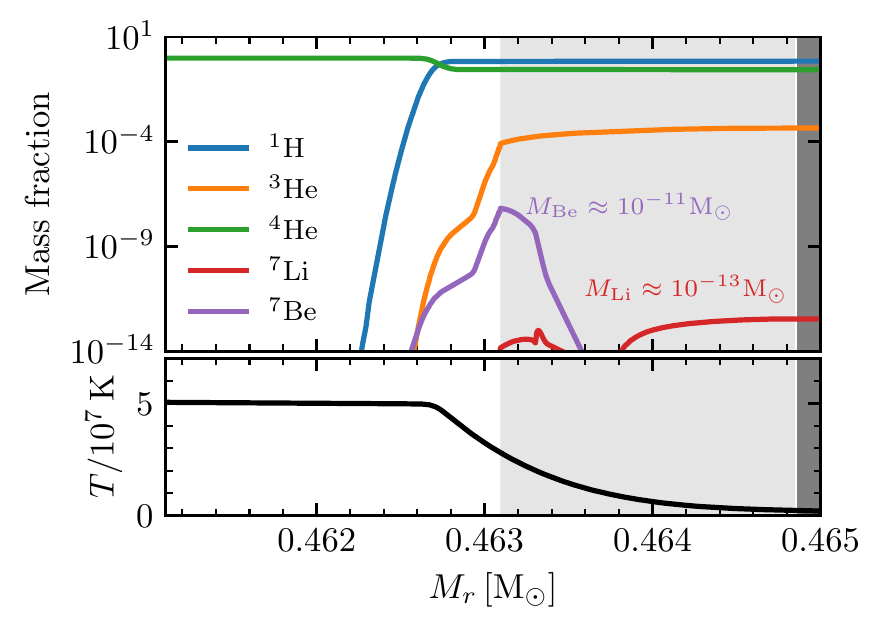}
  \caption{Properties of the $\unit[1]{\Msun}$ model in the vicinity
    of the H-burning shell just before the onset of the core He flash.
    The x-axis is the radial mass coordinate. The upper panel shows
    the mass fraction of key isotopes.  The total mass of \lithium[7]
    and \beryllium[7] present in the stellar model are indicated.  The
    lower panel shows the temperature.  The dark shaded region
    indicates the location of the envelope convective zone while the
    lighter shaded region shows areas experiencing thermohaline
    mixing.  Unshaded regions are radiative.}
  \label{fig:structure}
\end{figure}

In Figure~\ref{fig:structure}, we show the structure at the location
of the H-burning shell just before the core He flash.  The total mass
of \beryllium[7] present at this location is
$\approx \unit[10^{-11}]{\Msun}$ and is significantly larger than the
total mass of \lithium[7] in the star.  The structure of the H-burning
shell is set by the He core properties, which are approximately the
same in low-mass ($\lesssim \unit[2]{\Msun}$), near-core-flash stellar
models, and so the mass of \beryllium[7] should be approximately the
same in all low-mass RGs, independent of stellar mass.  If this
\beryllium[7]-rich region is mixed, without further synthesis or
destruction, into an $\approx \unit[0.5]{\Msun}$ convective envelope,
then this gives a characteristic abundance $\ALi \approx 0.6$,
approximately the abundance seen at the RC.  If the true \lithium[7]
production process is merely dilution and conversion of this pocket of
\beryllium[7], then since the stellar mass increases with the p-mode
large frequency separation, $\Delta \nu$, the observed surface lithium
abundance should decrease with increasing $\Delta \nu$.

\begin{figure}
  \centering
  \includegraphics[width=\columnwidth]{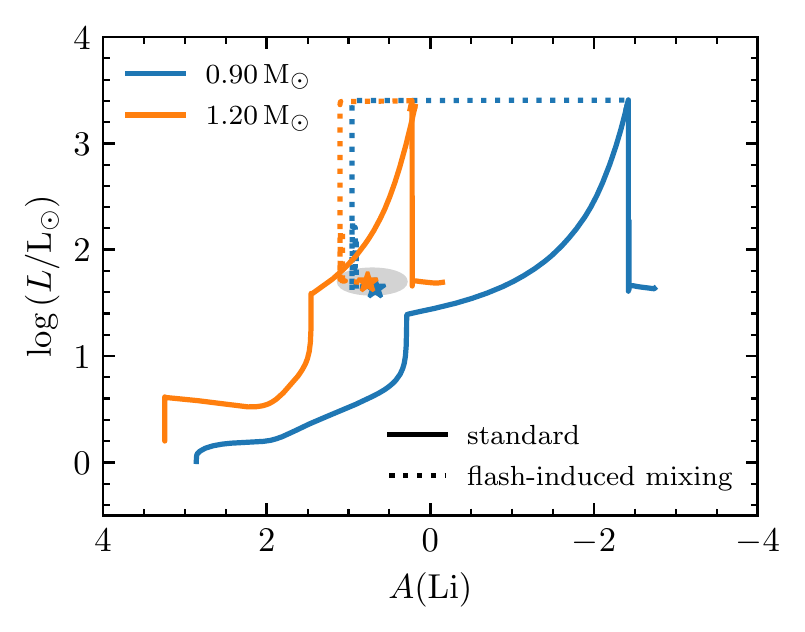}
  \caption{Same as Figure~\ref{fig:ALi}, but showing the evolutionary
    tracks for two different stellar masses.}
  \label{fig:mass}
\end{figure}

Figure~\ref{fig:mass} shows the evolution of $\unit[0.9]{\Msun}$ and
$\unit[1.2]{\Msun}$ models. All options except the initial mass are
unchanged.  Reproducing observations of the lithium abundances on the
main sequence and first ascent giant branch would require tuning the
mixing during these phases.  But despite the differences in lithium
abundances before the TRGB, because the structure above the H-burning
shell is similar, the flash-induced mixing yields similar abundances
on the RC.

A characteristic length scale to mix is
$\Delta R \sim \unit[10^{10}]{cm}$, as the width of the
\beryllium[7]-rich region is $\approx \unit[5\times10^9]{cm}$ and this
is located $\approx \unit[5\times10^{10}]{cm}$ below the lower
boundary of the convective envelope.
The mixing must occur on the timescale of the first sub-flash.
In our model $L_{\rm He} > 10^4\,\Lsun$ for $\approx \unit[100]{yr}$, 
so we take a characteristic duration for the sub-flash to be $\Delta t \sim \unit[10^9]{s}$.
This suggests a required effective mixing diffusion coefficient
$D_{\rm mix} \sim (\Delta R)^2/(\Delta t) \sim
\unit[10^{11}]{cm^2\,s^{-1}}$.

\begin{figure}
  \centering
  \includegraphics[width=\columnwidth]{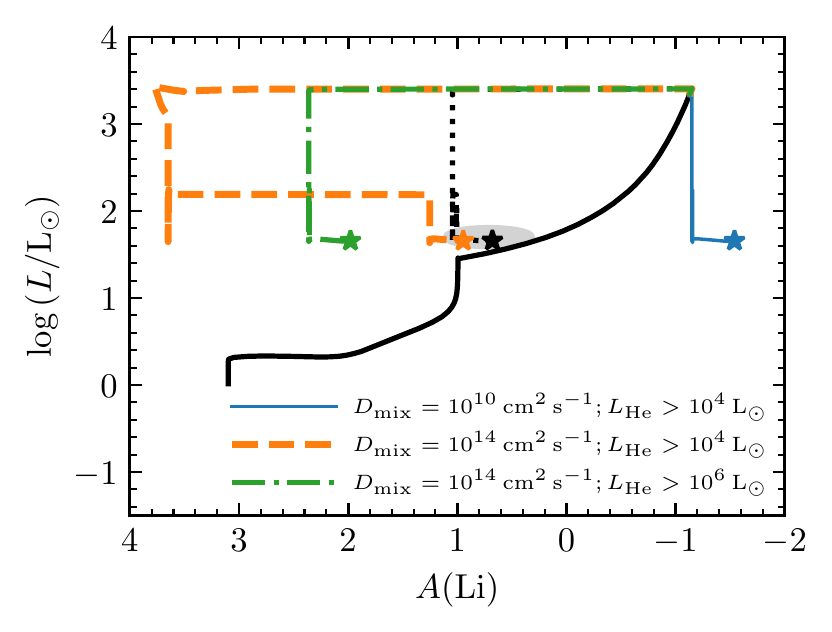}
  \caption{Effect of varying the strength and duration of the mixing
    event.  The flash-induced mixing case in Figure~\ref{fig:ALi},
    shown again as the black line, corresponds to
    $D_{\rm mix} = \unit[10^{12}]{cm^2\,s^{-1}}; L_{\rm He} >
    \unit[10^4]{\Lsun}$. }
  \label{fig:Dmix}
\end{figure}

Figure~\ref{fig:Dmix} shows tracks for different values of
$D_{\rm mix}$ and the threshold luminosity above which this mixing is
active.  Consistent with the estimate in the previous paragraph, a
value of $D_{\rm mix} = \unit[10^{10}]{cm^2\,s^{-1}}$ leads to a track
like the standard case in Figure~\ref{fig:ALi} (i.e., without a
flash-induced mixing event).  Increasing the strength of the mixing
allows for the efficient operation of the Cameron-Fowler process and
\lithium[7] synthesis well beyond the initial amount of \beryllium[7],
as seen in the track for
$D_{\rm mix} = \unit[10^{14}]{cm^2\,s^{-1}}; L_{\rm He} >
10^4\,\Lsun$. (We note, however, that this particular model then
experiences a lithium-depletion event associated with the second He
sub-flash on the way to the RC.)  Keeping this same higher mixing
value, but decreasing the duration of the mixing by increasing the
luminosity threshold to $L_{\rm He} > 10^6\,\Lsun$ (corresponding to a
duration of $\approx 4$ yr), results in an intermediate amount of
\lithium[7] synthesis.

\section{Estimates of IGW Mixing}

\begin{figure}
  \centering
  \includegraphics[width=\columnwidth]{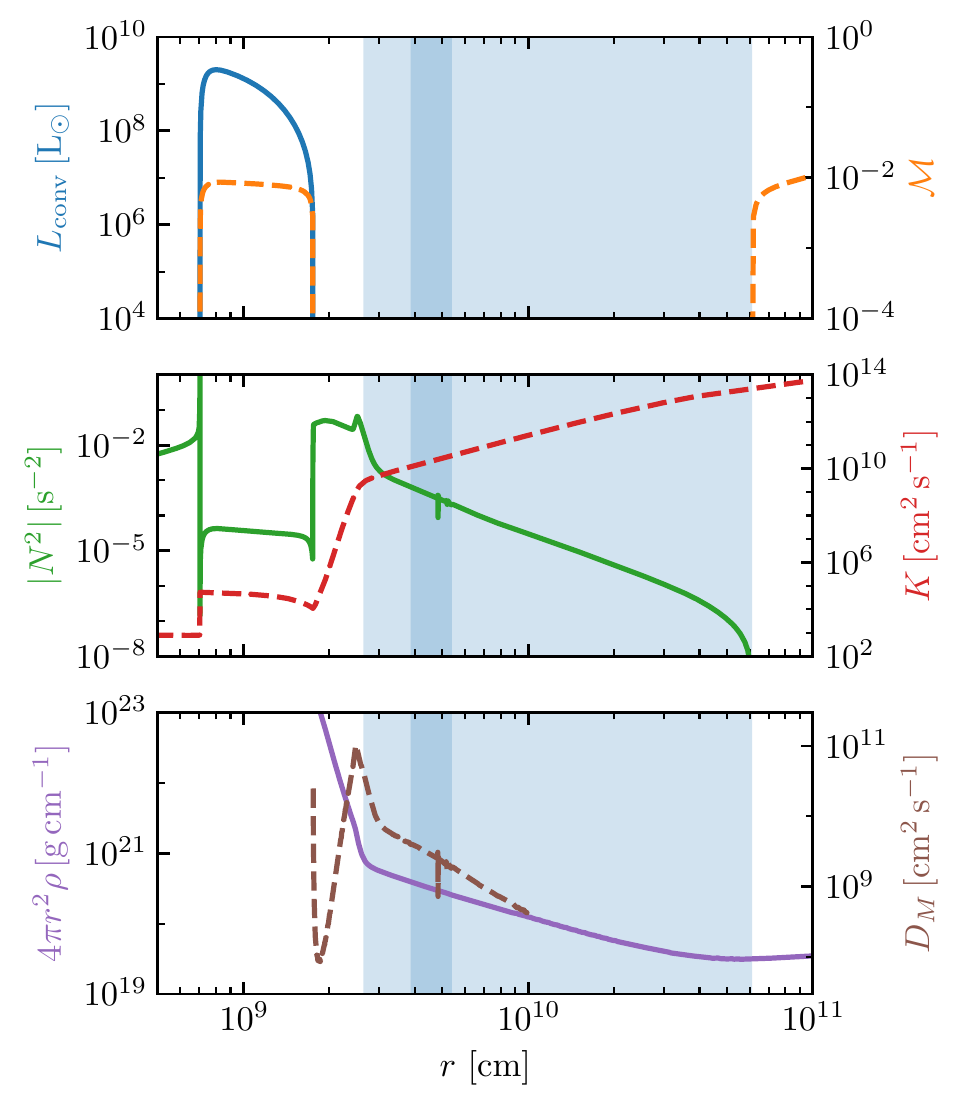}
  \caption{Overview of the stellar structure around the time of the peak luminosity of the first He
    sub-flash.  Solid lines correspond to the quantities on the left
    y-axes, and dashed lines correspond to quantities on the right
    y-axes.  The light shaded area indicates the region where we would
    impose the ad hoc mixing.  The darker area indicates the location
    of the \beryllium[7] pocket.}
  \label{fig:flash-structure}
\end{figure}

The ad hoc diffusivities invoked in the previous section illustrate
what is required of any mechanism that operates in rough concert with
the first He sub-flash.  Here, we provide some simple,
order-of-magnitude estimates to evaluate the potential effectiveness
of mixing from IGWs.

Figure~\ref{fig:flash-structure} shows the internal structure of
relevant parts of the star around the time of the peak luminosity of the first He
sub-flash.  We discuss each quantity as it enters our estimates, but
in all panels the light shaded region indicates where we would impose
our ad hoc mixing and the darker shaded region is the location of the
\beryllium[7].  As in Figure~\ref{fig:structure}, the mixing
region contains a mass of $\approx \unit[2 \times 10^{-3}]{\Msun}$.

\citet{Press1981} makes the energetic estimate that mixing a region
requires overcoming the gravitational potential energy and renewing
this energy each thermal diffusion timescale.  This corresponds to a
specific power $\sim N^2 K$, where $N$ is the Brunt-V\"ais\"al\"a
frequency and $K$ is the thermal diffusivity.  In the mixing region,
this is $\sim \unit[10^7]{erg\,s^{-1}\,g^{-1}}$, implying that the
luminosity required to induce mixing is
$L_{\rm mix} \sim \unit[10^4]{\Lsun}$.  This quantity is comparable to
the luminosity already transported radiatively in this region (i.e.,
$L_{\rm H}$).  At the time shown in the figure,
$L_{\rm IGW} \sim \mathcal{M} L_{\rm conv} \sim \unit[10^7]{\Lsun}$
and the condition $L_{\rm IGW} \gtrsim L_{\rm mix}$ is only satisfied
during the first sub-flash.  In the \citet{Press1981} framework, the
mass diffusivity associated with wave mixing has an upper limit at the
thermal diffusivity.  The ad hoc mixing invoked in the previous
section is of order the thermal diffusivity in the mixing region,
$K \sim \unit[10^9 - 10^{13}]{cm^2\,s^{-1}}$, implying that the
efficient wave mixing is required.

The thermal diffusion time across the mixing region is comparable to
the duration of the first sub-flash.  Therefore, an integrated energy
argument may be better suited to this case of a time-variable
luminosity.
The overall IGW energy budget is roughly
$\int dt\,\mathcal{M} L_{\rm He} \sim \unit[3\times10^{5}]{\Lsun\,yr}
\approx \unit[3\times10^{46}]{erg}$, where the integral is completely
dominated by the first sub-flash.  The energy required to mix this
stably-stratified region can be estimated as
$\int dm H_{P}^2 N^2 \sim \unit[10^{45}]{erg}$.  Therefore, this
mixing is energetically plausible, though it does require tapping a
significant fraction of the IGW energy.

Examining the profile of $N^2$ in the convective region, we see the
characteristic convective frequency is
$\omega^2 \sim \unit[3\times10^{-5}]{s^{-2}}$.  We expect
these waves to have a characteristic scale of the largest eddies in
the convective region (i.e., the pressure scale height), and so these
are $\ell \approx$ few modes with horizontal wavenumber,
$k_H = \sqrt{\ell(\ell+1)}/r \sim 1/r$.  The traveling waves can break---and hence lead to mixing as the wave energy is transferred into turbulence---if, as given in \citet{Lecoanet2019}, 
\begin{equation}
  L_{\rm IGW} \gtrsim %L_{\rm wave, break} \equiv
  \frac{4 \pi r^2 \rho \omega^4}{N k_{H}^3} \sim \unit[10^8]{\Lsun}~.
\end{equation}
(When we quote a single number estimate of such spatially-varying
quantities, it is the characteristic value evaluated in the
\beryllium[7] region.)  This estimate exceeds the peak value of
$L_{\rm IGW}$ implying that the traveling waves do not become
non-linear and break.

When IGW mixing is considered in the Sun, typically $\omega^2 \ll N^2$
and $N^2 \approx$ constant \citep[e.g.,][]{Montalban1994}.  The
profile of $N^2$ shows that neither of these conditions are true in
this situation.  Waves with the characteristic convective frequency
have a turning point in the mixing region.  The implies a reduction in
the wave luminosity with increasing radius since more of the wave
spectrum becomes unable to propagate as gravity waves as $N^2$
declines.  However, waves that reflect can continue to retain their
energy.
The thermal damping length for the waves \citep[e.g.,][]{Zahn1997} is
\begin{equation}
  L_{\rm damp} = \frac{4 \pi r^2 \rho \omega^4}{N N_{\rm T}^2 K} \sim \unit[10^{14}]{cm}~,
\end{equation}
where $N_T$ is the thermal part of the Brunt-V\"ais\"al\"a frequency.
(In the ad hoc mixing region, which is beyond the composition gradient
of the H-shell, $N_T \approx N$.)  The damping length is much longer
than the size of the cavity, so wave energy can accumulate in the
cavity and the increased amplitudes associated with standing waves in
the cavity may make non-linear effects more important.

To estimate the diffusivity from mixing by IGWs we follow the approach
of \citet{Montalban1994} and \citet{Montalban1996}, who find
\begin{equation}
  D_{M} = A^2 K^2 \left(\frac{N}{2\pi}\right) \left(\frac{\rho_0}{\rho}\right)^2  \left(\frac{r_0}{r}\right)^{12} f^{-2} u_M^5 \alpha_M^{-1}~,
\label{eq:DM}
\end{equation}
where
\begin{equation}
  f = \int_{r_0}^{r}K N^3 \left(\frac{r_0}{r'}\right)^{3} dr'
\end{equation}
and quantities with 0 subscripts are evaluated at the convective
boundary.  We assume the leading numerical constant $A^2 \sim 1$ and
adopt a characteristic horizontal eddy scale
$\alpha_M \sim \unit[10^9]{cm}$.  We follow their estimate for the
characteristic eddy velocity of
\begin{equation}
  u_M \sim \left(\frac{5 L_0}{4 \pi r_0^2 \rho_0}\right)^{1/3} \sim \unit[10^6]{cm\,s^{-1}},
\end{equation}
evaluated with $L_0 \sim \unit[10^7]{\Lsun}$.

We plot the value of Equation~\ref{eq:DM} in the bottom panel of
Figure~\ref{fig:flash-structure}.  Because the \citet{Montalban1994}
approach assumed $N^2 \gg \omega^2$, it cannot describe the entire ad
hoc mixing region, and we show it only where $N > \omega$.  At the
location of the \beryllium[7],
$D_M \sim \unit[3\times10^{9}]{cm^2\,s^{-1}}$.
This illustrates the plausibility of mixing of the required magnitude
from IGWs, but more detailed work specialized to this scenario will be
required to evaluate whether IGW-induced mixing can quantitatively
explain the observed phenomenon.

\section{Conclusions}

We have demonstrated that a brief chemical mixing event in the stellar
envelope, triggered at the same time as the first, most vigorous He
sub-flash, can reproduce the observed \lithium[7] enhancement of stars
on the red clump relative to the first ascent red giant branch.
We showed that the amount of \beryllium[7] present above the H-burning
shell just before the flash is sufficient, once mixed into the
envelope and converted to \lithium[7], to explain the typical
abundance on the red clump.
Given the large luminosities transported by convection in the core
during the helium flash, we suggest that the excitation of internal
gravity waves by the turbulent convection can provide a physical
mechanism to induce this mixing.
Our simple, ad hoc mixing prescription has limited predictive power,
but characterizes the required properties of the mixing.
The association with the first sub-flash does predict that objects
should be enhanced in lithium for their entire evolution between the
tip of the red giant branch and the red clump.
Future work using physical models of wave-induced mixing can determine
whether this process can meet the necessary criteria in terms of
mixing efficiency and spatial location.

\acknowledgments

We are grateful to Lars Bildsten, Jim Fuller, Adam Jermyn and Eliot Quataert for helpful discussions
and comments on the manuscript.
We thank the referee for suggestions that strengthened the presentation of this material.
JS is supported by the A.F. Morrison Fellowship in Lick Observatory.
We acknowledge use of the lux supercomputer at UC Santa Cruz, funded by NSF MRI grant AST 1828315.

\software{
\MESA\ \citep{Paxton2011, Paxton2013, Paxton2015, Paxton2018, Paxton2019},
\texttt{MESASDK} 20.3.1 \citep{mesasdk}, 
\texttt{matplotlib} \citep{hunter2007}, 
\texttt{NumPy} \citep{walt2011},
\texttt{py\_mesa\_reader} \citep{pmr},
\texttt{MesaScript} \citep{MesaScript}
}

\clearpage

\bibliography{Li}

\begin{thebibliography}{}
\expandafter\ifx\csname natexlab\endcsname\relax\def\natexlab#1{#1}\fi
\providecommand{\url}[1]{\href{#1}{#1}}
\providecommand{\dodoi}[1]{doi:~\href{http://doi.org/#1}{\nolinkurl{#1}}}
\providecommand{\doeprint}[1]{\href{http://ascl.net/#1}{\nolinkurl{http://ascl.net/#1}}}
\providecommand{\doarXiv}[1]{\href{https://arxiv.org/abs/#1}{\nolinkurl{https://arxiv.org/abs/#1}}}

\bibitem[{{Angelou} {et~al.}(2012){Angelou}, {Stancliffe}, {Church},
  {Lattanzio}, \& {Smith}}]{Angelou2012}
{Angelou}, G.~C., {Stancliffe}, R.~J., {Church}, R.~P., {Lattanzio}, J.~C., \&
  {Smith}, G.~H. 2012, \apj, 749, 128, \dodoi{10.1088/0004-637X/749/2/128}

\bibitem[{{Angulo} {et~al.}(1999){Angulo}, {Arnould}, {Rayet}, {Descouvemont},
  {Baye}, {Leclercq-Willain}, {Coc}, {Barhoumi}, {Aguer}, {Rolfs}, {Kunz},
  {Hammer}, {Mayer}, {Paradellis}, {Kossionides}, {Chronidou}, {Spyrou},
  {degl'Innocenti}, {Fiorentini}, {Ricci}, {Zavatarelli}, {Providencia},
  {Wolters}, {Soares}, {Grama}, {Rahighi}, {Shotter}, \& {Lamehi
  Rachti}}]{Angulo1999}
{Angulo}, C., {Arnould}, M., {Rayet}, M., {et~al.} 1999, Nuclear Physics A,
  656, 3, \dodoi{10.1016/S0375-9474(99)00030-5}

\bibitem[{{Asplund} {et~al.}(2009){Asplund}, {Grevesse}, {Sauval}, \&
  {Scott}}]{Asplund2009}
{Asplund}, M., {Grevesse}, N., {Sauval}, A.~J., \& {Scott}, P. 2009, \araa, 47,
  481, \dodoi{10.1146/annurev.astro.46.060407.145222}

\bibitem[{{Baraffe} {et~al.}(2017){Baraffe}, {Pratt}, {Goffrey}, {Constantino},
  {Folini}, {Popov}, {Walder}, \& {Viallet}}]{Baraffe2017b}
{Baraffe}, I., {Pratt}, J., {Goffrey}, T., {et~al.} 2017, \apjl, 845, L6,
  \dodoi{10.3847/2041-8213/aa82ff}

\bibitem[{{Bedding} {et~al.}(2011){Bedding}, {Mosser}, {Huber},
  {Montalb{\'a}n}, {Beck}, {Christensen-Dalsgaard}, {Elsworth},
  {Garc{\'{\i}}a}, {Miglio}, {Stello}, {White}, {De Ridder}, {Hekker}, {Aerts},
  {Barban}, {Belkacem}, {Broomhall}, {Brown}, {Buzasi}, {Carrier}, {Chaplin},
  {di Mauro}, {Dupret}, {Frandsen}, {Gilliland}, {Goupil}, {Jenkins},
  {Kallinger}, {Kawaler}, {Kjeldsen}, {Mathur}, {Noels}, {Aguirre}, \&
  {Ventura}}]{Bedding2011}
{Bedding}, T.~R., {Mosser}, B., {Huber}, D., {et~al.} 2011, \nat, 471, 608,
  \dodoi{10.1038/nature09935}

\bibitem[{{Bildsten} {et~al.}(2012){Bildsten}, {Paxton}, {Moore}, \&
  {Macias}}]{Bildsten2012}
{Bildsten}, L., {Paxton}, B., {Moore}, K., \& {Macias}, P.~J. 2012, \apjl, 744,
  L6, \dodoi{10.1088/2041-8205/744/1/L6}

\bibitem[{{Buder} {et~al.}(2018){Buder}, {Asplund}, {Duong}, {Kos}, {Lind},
  {Ness}, {Sharma}, {Bland -Hawthorn}, {Casey}, {de Silva}, {D'Orazi},
  {Freeman}, {Lewis}, {Lin}, {Martell}, {Schlesinger}, {Simpson}, {Zucker},
  {Zwitter}, {Amarsi}, {Anguiano}, {Carollo}, {Casagrande}, {{\v{C}}otar},
  {Cottrell}, {da Costa}, {Gao}, {Hayden}, {Horner}, {Ireland}, {Kafle},
  {Munari}, {Nataf}, {Nordlander}, {Stello}, {Ting}, {Traven}, {Watson},
  {Wittenmyer}, {Wyse}, {Yong}, {Zinn}, {{\v{Z}}erjal}, \& {Galah
  Collaboration}}]{GALAHDR2}
{Buder}, S., {Asplund}, M., {Duong}, L., {et~al.} 2018, \mnras, 478, 4513,
  \dodoi{10.1093/mnras/sty1281}

\bibitem[{{Cameron}(1955)}]{Cameron1955}
{Cameron}, A.~G.~W. 1955, \apj, 121, 144, \dodoi{10.1086/145970}

\bibitem[{{Cameron} \& {Fowler}(1971)}]{Cameron1971}
{Cameron}, A.~G.~W., \& {Fowler}, W.~A. 1971, \apj, 164, 111,
  \dodoi{10.1086/150821}

\bibitem[{{Charbonnel} \& {Lagarde}(2010)}]{Charbonnel2010}
{Charbonnel}, C., \& {Lagarde}, N. 2010, \aap, 522, A10,
  \dodoi{10.1051/0004-6361/201014432}

\bibitem[{{Charbonnel} \& {Talon}(2005)}]{Charbonnel2005}
{Charbonnel}, C., \& {Talon}, S. 2005, Science, 309, 2189,
  \dodoi{10.1126/science.1116849}

\bibitem[{{Charbonnel} \& {Zahn}(2007)}]{Charbonnel2007}
{Charbonnel}, C., \& {Zahn}, J.-P. 2007, \aap, 467, L15,
  \dodoi{10.1051/0004-6361:20077274}

\bibitem[{{Cyburt} {et~al.}(2010){Cyburt}, {Amthor}, {Ferguson}, {Meisel},
  {Smith}, {Warren}, {Heger}, {Hoffman}, {Rauscher}, {Sakharuk}, {Schatz},
  {Thielemann}, \& {Wiescher}}]{Cyburt2010}
{Cyburt}, R.~H., {Amthor}, A.~M., {Ferguson}, R., {et~al.} 2010, \apjs, 189,
  240, \dodoi{10.1088/0067-0049/189/1/240}

\bibitem[{{Gaia Collaboration} {et~al.}(2018){Gaia Collaboration}, {Brown},
  {Vallenari}, {Prusti}, {de Bruijne}, {Babusiaux}, {Bailer-Jones}, {Biermann},
  {Evans}, {Eyer}, {Jansen}, {Jordi}, {Klioner}, {Lammers}, {Lindegren},
  {Luri}, {Mignard}, {Panem}, {Pourbaix}, {Randich}, {Sartoretti}, {Siddiqui},
  {Soubiran}, {van Leeuwen}, {Walton}, {Arenou}, {Bastian}, {Cropper},
  {Drimmel}, {Katz}, {Lattanzi}, {Bakker}, {Cacciari}, {Casta{\~n}eda},
  {Chaoul}, {Cheek}, {De Angeli}, {Fabricius}, {Guerra}, {Holl}, {Masana},
  {Messineo}, {Mowlavi}, {Nienartowicz}, {Panuzzo}, {Portell}, {Riello},
  {Seabroke}, {Tanga}, {Th{\'e}venin}, {Gracia-Abril}, {Comoretto},
  {Garcia-Reinaldos}, {Teyssier}, {Altmann}, {Andrae}, {Audard},
  {Bellas-Velidis}, {Benson}, {Berthier}, {Blomme}, {Burgess}, {Busso},
  {Carry}, {Cellino}, {Clementini}, {Clotet}, {Creevey}, {Davidson}, {De
  Ridder}, {Delchambre}, {Dell'Oro}, {Ducourant},
  {Fern{\'a}ndez-Hern{\'a}ndez}, {Fouesneau}, {Fr{\'e}mat}, {Galluccio},
  {Garc{\'\i}a-Torres}, {Gonz{\'a}lez-N{\'u}{\~n}ez}, {Gonz{\'a}lez-Vidal},
  {Gosset}, {Guy}, {Halbwachs}, {Hambly}, {Harrison}, {Hern{\'a}ndez},
  {Hestroffer}, {Hodgkin}, {Hutton}, {Jasniewicz}, {Jean-Antoine-Piccolo},
  {Jordan}, {Korn}, {Krone-Martins}, {Lanzafame}, {Lebzelter}, {L{\"o}ffler},
  {Manteiga}, {Marrese}, {Mart{\'\i}n-Fleitas}, {Moitinho}, {Mora}, {Muinonen},
  {Osinde}, {Pancino}, {Pauwels}, {Petit}, {Recio-Blanco}, {Richards},
  {Rimoldini}, {Robin}, {Sarro}, {Siopis}, {Smith}, {Sozzetti}, {S{\"u}veges},
  {Torra}, {van Reeven}, {Abbas}, {Abreu Aramburu}, {Accart}, {Aerts},
  {Altavilla}, {{\'A}lvarez}, {Alvarez}, {Alves}, {Anderson}, {Andrei},
  {Anglada Varela}, {Antiche}, {Antoja}, {Arcay}, {Astraatmadja}, {Bach},
  {Baker}, {Balaguer-N{\'u}{\~n}ez}, {Balm}, {Barache}, {Barata}, {Barbato},
  {Barblan}, {Barklem}, {Barrado}, {Barros}, {Barstow}, {Bartholom{\'e}
  Mu{\~n}oz}, {Bassilana}, {Becciani}, {Bellazzini}, {Berihuete}, {Bertone},
  {Bianchi}, {Bienaym{\'e}}, {Blanco-Cuaresma}, {Boch}, {Boeche}, {Bombrun},
  {Borrachero}, {Bossini}, {Bouquillon}, {Bourda}, {Bragaglia}, {Bramante},
  {Breddels}, {Bressan}, {Brouillet}, {Br{\"u}semeister}, {Brugaletta},
  {Bucciarelli}, {Burlacu}, {Busonero}, {Butkevich}, {Buzzi}, {Caffau},
  {Cancelliere}, {Cannizzaro}, {Cantat-Gaudin}, {Carballo}, {Carlucci},
  {Carrasco}, {Casamiquela}, {Castellani}, {Castro-Ginard}, {Charlot},
  {Chemin}, {Chiavassa}, {Cocozza}, {Costigan}, {Cowell}, {Crifo}, {Crosta},
  {Crowley}, {Cuypers}, {Dafonte}, {Damerdji}, {Dapergolas}, {David}, {David},
  {de Laverny}, {De Luise}, {De March}, {de Martino}, {de Souza}, {de Torres},
  {Debosscher}, {del Pozo}, {Delbo}, {Delgado}, {Delgado}, {Di Matteo},
  {Diakite}, {Diener}, {Distefano}, {Dolding}, {Drazinos}, {Dur{\'a}n},
  {Edvardsson}, {Enke}, {Eriksson}, {Esquej}, {Eynard Bontemps}, {Fabre},
  {Fabrizio}, {Faigler}, {Falc{\~a}o}, {Farr{\`a}s Casas}, {Federici},
  {Fedorets}, {Fernique}, {Figueras}, {Filippi}, {Findeisen}, {Fonti},
  {Fraile}, {Fraser}, {Fr{\'e}zouls}, {Gai}, {Galleti}, {Garabato},
  {Garc{\'\i}a-Sedano}, {Garofalo}, {Garralda}, {Gavel}, {Gavras}, {Gerssen},
  {Geyer}, {Giacobbe}, {Gilmore}, {Girona}, {Giuffrida}, {Glass}, {Gomes},
  {Granvik}, {Gueguen}, {Guerrier}, {Guiraud}, {Guti{\'e}rrez-S{\'a}nchez},
  {Haigron}, {Hatzidimitriou}, {Hauser}, {Haywood}, {Heiter}, {Helmi}, {Heu},
  {Hilger}, {Hobbs}, {Hofmann}, {Holland}, {Huckle}, {Hypki}, {Icardi},
  {Jan{\ss}en}, {Jevardat de Fombelle}, {Jonker}, {Juh{\'a}sz}, {Julbe},
  {Karampelas}, {Kewley}, {Klar}, {Kochoska}, {Kohley}, {Kolenberg},
  {Kontizas}, {Kontizas}, {Koposov}, {Kordopatis}, {Kostrzewa-Rutkowska},
  {Koubsky}, {Lambert}, {Lanza}, {Lasne}, {Lavigne}, {Le Fustec}, {Le
  Poncin-Lafitte}, {Lebreton}, {Leccia}, {Leclerc}, {Lecoeur-Taibi},
  {Lenhardt}, {Leroux}, {Liao}, {Licata}, {Lindstr{\o}m}, {Lister}, {Livanou},
  {Lobel}, {L{\'o}pez}, {Managau}, {Mann}, {Mantelet}, {Marchal}, {Marchant},
  {Marconi}, {Marinoni}, {Marschalk{\'o}}, {Marshall}, {Martino}, {Marton},
  {Mary}, {Massari}, {Matijevi{\v{c}}}, {Mazeh}, {McMillan}, {Messina},
  {Michalik}, {Millar}, {Molina}, {Molinaro}, {Moln{\'a}r}, {Montegriffo},
  {Mor}, {Morbidelli}, {Morel}, {Morris}, {Mulone}, {Muraveva}, {Musella},
  {Nelemans}, {Nicastro}, {Noval}, {O'Mullane}, {Ord{\'e}novic},
  {Ord{\'o}{\~n}ez-Blanco}, {Osborne}, {Pagani}, {Pagano}, {Pailler},
  {Palacin}, {Palaversa}, {Panahi}, {Pawlak}, {Piersimoni}, {Pineau}, {Plachy},
  {Plum}, {Poggio}, {Poujoulet}, {Pr{\v{s}}a}, {Pulone}, {Racero}, {Ragaini},
  {Rambaux}, {Ramos-Lerate}, {Regibo}, {Reyl{\'e}}, {Riclet}, {Ripepi}, {Riva},
  {Rivard}, {Rixon}, {Roegiers}, {Roelens}, {Romero-G{\'o}mez}, {Rowell},
  {Royer}, {Ruiz-Dern}, {Sadowski}, {Sagrist{\`a} Sell{\'e}s}, {Sahlmann},
  {Salgado}, {Salguero}, {Sanna}, {Santana-Ros}, {Sarasso}, {Savietto},
  {Schultheis}, {Sciacca}, {Segol}, {Segovia}, {S{\'e}gransan}, {Shih},
  {Siltala}, {Silva}, {Smart}, {Smith}, {Solano}, {Solitro}, {Sordo}, {Soria
  Nieto}, {Souchay}, {Spagna}, {Spoto}, {Stampa}, {Steele},
  {Steidelm{\"u}ller}, {Stephenson}, {Stoev}, {Suess}, {Surdej}, {Szabados},
  {Szegedi-Elek}, {Tapiador}, {Taris}, {Tauran}, {Taylor}, {Teixeira},
  {Terrett}, {Teyssand ier}, {Thuillot}, {Titarenko}, {Torra Clotet}, {Turon},
  {Ulla}, {Utrilla}, {Uzzi}, {Vaillant}, {Valentini}, {Valette}, {van Elteren},
  {Van Hemelryck}, {van Leeuwen}, {Vaschetto}, {Vecchiato}, {Veljanoski},
  {Viala}, {Vicente}, {Vogt}, {von Essen}, {Voss}, {Votruba}, {Voutsinas},
  {Walmsley}, {Weiler}, {Wertz}, {Wevers}, {Wyrzykowski}, {Yoldas},
  {{\v{Z}}erjal}, {Ziaeepour}, {Zorec}, {Zschocke}, {Zucker}, {Zurbach}, \&
  {Zwitter}}]{GaiaDR2}
{Gaia Collaboration}, {Brown}, A.~G.~A., {Vallenari}, A., {et~al.} 2018, \aap,
  616, A1, \dodoi{10.1051/0004-6361/201833051}

\bibitem[{{Garcia Lopez} \& {Spruit}(1991)}]{GarciaLopez1991}
{Garcia Lopez}, R.~J., \& {Spruit}, H.~C. 1991, \apj, 377, 268,
  \dodoi{10.1086/170356}

\bibitem[{{Goldreich} \& {Kumar}(1990)}]{Goldreich1990}
{Goldreich}, P., \& {Kumar}, P. 1990, \apj, 363, 694, \dodoi{10.1086/169376}

\bibitem[{{Hawkins} {et~al.}(2018){Hawkins}, {Ting}, \&
  {Walter-Rix}}]{Hawkins2018a}
{Hawkins}, K., {Ting}, Y.-S., \& {Walter-Rix}, H. 2018, \apj, 853, 20,
  \dodoi{10.3847/1538-4357/aaa08a}

\bibitem[{Hunter(2007)}]{hunter2007}
Hunter, J.~D. 2007, Computing In Science \&amp; Engineering, 9, 90

\bibitem[{{Kippenhahn} {et~al.}(1980){Kippenhahn}, {Ruschenplatt}, \&
  {Thomas}}]{Kippenhahn1980a}
{Kippenhahn}, R., {Ruschenplatt}, G., \& {Thomas}, H.-C. 1980, \aap, 91, 175

\bibitem[{{Kumar} {et~al.}(2020){Kumar}, {Reddy}, {Campbell}, {Maben}, {Zhao},
  \& {Ting}}]{Kumar2020b}
{Kumar}, Y.~B., {Reddy}, B.~E., {Campbell}, S.~W., {et~al.} 2020, Nature
  Astronomy, \dodoi{10.1038/s41550-020-1139-7}

\bibitem[{{Lecoanet} \& {Quataert}(2013)}]{Lecoanet2013}
{Lecoanet}, D., \& {Quataert}, E. 2013, \mnras, 430, 2363,
  \dodoi{10.1093/mnras/stt055}

\bibitem[{{Lecoanet} {et~al.}(2019){Lecoanet}, {Cantiello}, {Quataert},
  {Couston}, {Burns}, {Pope}, {Jermyn}, {Favier}, \& {Le Bars}}]{Lecoanet2019}
{Lecoanet}, D., {Cantiello}, M., {Quataert}, E., {et~al.} 2019, \apjl, 886,
  L15, \dodoi{10.3847/2041-8213/ab5446}

\bibitem[{{Martell} {et~al.}(2020){Martell}, {Simpson}, {Balasubramaniam},
  {Buder}, {Sharma}, {Hon}, {Stello}, {Ting}, {Asplund}, {Bland -Hawthorn}, {De
  Silva}, {Freeman}, {Hayden}, {Kos}, {Lewis}, {Lind}, {Zucker}, {Zwitter},
  {Campbell}, {Cotar}, {Horner}, {Montet}, \& {Wittenmyer}}]{Martell2020}
{Martell}, S., {Simpson}, J., {Balasubramaniam}, A., {et~al.} 2020, arXiv
  e-prints, arXiv:2006.02106.
\newblock \doarXiv{2006.02106}

\bibitem[{{Miller Bertolami} {et~al.}(2020){Miller Bertolami}, {Battich},
  {C{\'o}rsico}, {Christensen-Dalsgaard}, \& {Althaus}}]{MillerBertolami2020}
{Miller Bertolami}, M.~M., {Battich}, T., {C{\'o}rsico}, A.~H.,
  {Christensen-Dalsgaard}, J., \& {Althaus}, L.~G. 2020, Nature Astronomy, 4,
  67, \dodoi{10.1038/s41550-019-0890-0}

\bibitem[{{Montalban}(1994)}]{Montalban1994}
{Montalban}, J. 1994, \aap, 281, 421

\bibitem[{{Montalban} \& {Schatzman}(1996)}]{Montalban1996}
{Montalban}, J., \& {Schatzman}, E. 1996, \aap, 305, 513

\bibitem[{{Mosser} {et~al.}(2011){Mosser}, {Barban}, {Montalb{\'a}n}, {Beck},
  {Miglio}, {Belkacem}, {Goupil}, {Hekker}, {De Ridder}, {Dupret}, {Elsworth},
  {Noels}, {Baudin}, {Michel}, {Samadi}, {Auvergne}, {Baglin}, \&
  {Catala}}]{Mosser2011}
{Mosser}, B., {Barban}, C., {Montalb{\'a}n}, J., {et~al.} 2011, \aap, 532, A86,
  \dodoi{10.1051/0004-6361/201116825}

\bibitem[{{Paxton} {et~al.}(2011){Paxton}, {Bildsten}, {Dotter}, {Herwig},
  {Lesaffre}, \& {Timmes}}]{Paxton2011}
{Paxton}, B., {Bildsten}, L., {Dotter}, A., {et~al.} 2011, \apjs, 192, 3,
  \dodoi{10.1088/0067-0049/192/1/3}

\bibitem[{{Paxton} {et~al.}(2013){Paxton}, {Cantiello}, {Arras}, {Bildsten},
  {Brown}, {Dotter}, {Mankovich}, {Montgomery}, {Stello}, {Timmes}, \&
  {Townsend}}]{Paxton2013}
{Paxton}, B., {Cantiello}, M., {Arras}, P., {et~al.} 2013, \apjs, 208, 4,
  \dodoi{10.1088/0067-0049/208/1/4}

\bibitem[{{Paxton} {et~al.}(2015){Paxton}, {Marchant}, {Schwab}, {Bauer},
  {Bildsten}, {Cantiello}, {Dessart}, {Farmer}, {Hu}, {Langer}, {Townsend},
  {Townsley}, \& {Timmes}}]{Paxton2015}
{Paxton}, B., {Marchant}, P., {Schwab}, J., {et~al.} 2015, \apjs, 220, 15,
  \dodoi{10.1088/0067-0049/220/1/15}

\bibitem[{{Paxton} {et~al.}(2018){Paxton}, {Schwab}, {Bauer}, {Bildsten},
  {Blinnikov}, {Duffell}, {Farmer}, {Goldberg}, {Marchant}, {Sorokina},
  {Thoul}, {Townsend}, \& {Timmes}}]{Paxton2018}
{Paxton}, B., {Schwab}, J., {Bauer}, E.~B., {et~al.} 2018, \apjs, 234, 34,
  \dodoi{10.3847/1538-4365/aaa5a8}

\bibitem[{{Paxton} {et~al.}(2019){Paxton}, {Smolec}, {Schwab}, {Gautschy},
  {Bildsten}, {Cantiello}, {Dotter}, {Farmer}, {Goldberg}, {Jermyn}, {Kanbur},
  {Marchant}, {Thoul}, {Townsend}, {Wolf}, {Zhang}, \& {Timmes}}]{Paxton2019}
{Paxton}, B., {Smolec}, R., {Schwab}, J., {et~al.} 2019, \apjs, 243, 10,
  \dodoi{10.3847/1538-4365/ab2241}

\bibitem[{{Press}(1981)}]{Press1981}
{Press}, W.~H. 1981, \apj, 245, 286, \dodoi{10.1086/158809}

\bibitem[{{Simonucci} {et~al.}(2013){Simonucci}, {Taioli}, {Palmerini}, \&
  {Busso}}]{Simonucci2013}
{Simonucci}, S., {Taioli}, S., {Palmerini}, S., \& {Busso}, M. 2013, \apj, 764,
  118, \dodoi{10.1088/0004-637X/764/2/118}

\bibitem[{{Thomas}(1967)}]{Thomas1967}
{Thomas}, H.-C. 1967, \zap, 67, 420

\bibitem[{{Ting} {et~al.}(2018){Ting}, {Hawkins}, \& {Rix}}]{Ting2018c}
{Ting}, Y.-S., {Hawkins}, K., \& {Rix}, H.-W. 2018, \apjl, 858, L7,
  \dodoi{10.3847/2041-8213/aabf8e}

\bibitem[{Townsend(2020)}]{mesasdk}
Townsend, R. 2020, MESA SDK for Linux, 20.3.1,  Zenodo,
  \dodoi{10.5281/zenodo.3706650}

\bibitem[{van~der Walt {et~al.}(2011)van~der Walt, Colbert, \&
  Varoquaux}]{walt2011}
van~der Walt, S., Colbert, S.~C., \& Varoquaux, G. 2011, Computing in Science
  Engineering, 13, 22, \dodoi{10.1109/MCSE.2011.37}

\bibitem[{{Vescovi} {et~al.}(2019){Vescovi}, {Piersanti}, {Cristallo}, {Busso},
  {Vissani}, {Palmerini}, {Simonucci}, \& {Taioli}}]{Vescovi2019}
{Vescovi}, D., {Piersanti}, L., {Cristallo}, S., {et~al.} 2019, \aap, 623,
  A126, \dodoi{10.1051/0004-6361/201834993}

\bibitem[{Wolf {et~al.}(2017)Wolf, Bauer, \& Schwab}]{MesaScript}
Wolf, B., Bauer, E.~B., \& Schwab, J. 2017, {wmwolf/MesaScript: A DSL for
  Writing MESA Inlists}, \dodoi{10.5281/zenodo.826954}

\bibitem[{Wolf \& Schwab(2017)}]{pmr}
Wolf, B., \& Schwab, J. 2017, wmwolf/py\_mesa\_reader: Interact with MESA
  Output, \dodoi{10.5281/zenodo.826958}

\bibitem[{{Zahn} {et~al.}(1997){Zahn}, {Talon}, \& {Matias}}]{Zahn1997}
{Zahn}, J.~P., {Talon}, S., \& {Matias}, J. 1997, \aap, 322, 320.
\newblock \doarXiv{astro-ph/9611189}

\end{thebibliography}

\end{document}